\documentclass[runningheads]{llncs}
\usepackage{graphicx}
\usepackage{amssymb}
\begin{document}
\title{A Novel Approach to Analyze Fashion Digital Archive from Humanities\thanks{This work was supported by the Japan Society for the Promotion of Science (JSPS) KAKENHI [grant number 20K02153].}}
%
%
\author{Satoshi Takahashi\inst{1}\orcidID{0000-0002-1067-6704}   \and
Keiko Yamaguchi\inst{2}\orcidID{0000-0002-1129-8012} \and
Asuka Watanabe\inst{3}\orcidID{0000-0002-3539-8253}}
\authorrunning{S. Takahashi et al.}
%
\institute{Kanto Gakuin University, Mutsuura, Kanazawa, Yokohama, Kanagawa, Japan \and
Nagoya University, Furocho, Chikusa, Nagoya, Aichi, Japan \and
Kyoritsu Women’s Junior College, 2-6-1 Hitotsubashi, Chiyoda, Tokyo, Japan
\email{satotaka@kanto-gakuin.ac.jp}\\
}
\maketitle              
\begin{abstract}
Fashion styles adopted every day are an important aspect of culture, and style trend analysis helps provide a deeper understanding of our societies and cultures. To analyze everyday fashion trends from the humanities perspective, we need a digital archive that includes images of what people wore in their daily lives over an extended period. In fashion research, building digital fashion image archives has attracted significant attention. However, the existing archives are not suitable for retrieving everyday fashion trends. In addition, to interpret how the trends emerge, we need non-fashion data sources relevant to why and how people choose fashion. In this study, we created a new fashion image archive called Chronicle Archive of Tokyo Street Fashion (CAT STREET) based on a review of the limitations in the existing digital fashion archives. CAT STREET includes images showing the clothing people wore in their daily lives during the period 1970--2017, which contain timestamps and street location annotations. We applied machine learning to CAT STREET and found two types of fashion trend patterns. Then, we demonstrated how magazine archives help us interpret how trend patterns emerge. These empirical analyses show our approach’s potential to discover new perspectives to promote an understanding of our societies and cultures through fashion embedded in consumers’ daily lives.

\keywords{Fashion trend \and Digital fashion archive \and Image processing \and Machine learning \and Deep learning.}
\end{abstract}
\section{Introduction}
Fashion styles adopted in our daily lives are an important aspect of culture. As noted by Lancioni~\cite{lancioni1973brief}, fashion is ‘a reflection of a society’s goals and aspirations’; people choose fashion styles within their embedded social contexts. Hence, the analysis of everyday fashion trends can provide an in-depth understanding of our societies and cultures. In fashion research, the advantages of digital fashion archives have continued to attract attention in recent times. These digital archives have been mainly built based on two purposes. One is enhancing the scholarly values of museum collections by making them public to utilize them digitally. Another motivation is predicting what fashion items will come into style in the short term with images retrieved from the internet for business purposes.

However, these archives are not suitable for analyzing everyday fashion trends from the humanities perspective. To identify everyday fashion trends, we need a new digital archive that includes images of the clothes people wore in their daily lives over an extended period. In addition, we need non-fashion data sources showcasing why and how people choose fashion to interpret the trends.

In this study, we create a new fashion image archive called Chronicle Archive of Tokyo Street Fashion (CAT STREET) showcasing fashion images reflecting everyday women’s clothing over a long period from 1970 to 2017 in Tokyo. CAT STREET helps overcome the limitations in the existing digital fashion archives by applying machine learning to identify fashion trend patterns. Then, we demonstrate how magazine archives are suitable in understanding how various trend patterns emerged. The empirical analyses show our approach’s potential in identifying new perspectives through fashion trends in daily life, which can promote understanding of societies and cultures.

\section{Related Works}
\subsection{Fashion Trend Analysis}
Kroeber~\cite{kroeber1919principle} analyzed fashion trends manually and measured features of women’s full evening toilette, e.g. skirt length and skirt width, which were collected from magazines from 1844 to 1919. After this work, many researchers conducted similar studies~\cite{belleau1987cyclical,lowe1993quantitative,lowe1990velocity,robenstine1981relating}. Their work surveyed historical magazines, portraits, and illustrations and focused mainly on formal fashion, rather than clothing worn in everyday life.

The critical issue in the traditional approach is that analyzing fashion trends is labor-intensive. Kroeber~\cite{kroeber1919principle} is among the representative works in the early stage of quantitative analysis of fashion trends. This type of quantitative analysis requires a considerable amount of human resources to select appropriate images, classify images, and measure features. This labor-intensive approach has long been applied in this field. Furthermore, manually managing large modern fashion image archives is cumbersome. To solve this issue, we utilize machine learning in this study. Machine learning is a computer algorithm that learns procedures based on sample data, e.g. human task results, and imitates the procedures. In recent years, machine learning has contributed to the development of digital humanities information processing~\cite{kestemont2017lemmatization,lang2018attesting,wevers2020visual}. In the field of fashion, modern fashion styles are complex and diverse. Applying machine learning to fashion image archives may be expected to be of benefit in quantifying fashion trends more precisely and efficiently.

\subsection{Digital Fashion Archives}
In recent years, digital fashion archives have been proactively built in the fields of museology and computer science. However, the construction of these archives was prompted by different research objectives. Many museums and research institutions have digitized their fashion collections~\cite{Vogue,EuropeanaFashion,JapaneseFashionArchive,LACMAFashion,BostonMuseum,MET,TheMuseumatFIT,UNT} not only to conserve these historically valuable collections but also to enhance the scholarly values by making them available to the public. Vogue digitized their magazines from 1892 to date, the Japanese Fashion Archive collected representative Japanese fashion items from each era, and the Boston Museum of Fine Arts archived historical fashion items from all over the world. 

On the other hand, in computer science, the motivation to build digital fashion archives is generally short-sighted and primarily to predict the next fashion trends that can be used by vendors to plan production or by online stores to improve recommendation engines. Several studies have created their own fashion image databases to fit the fashion business issues they focused on~\cite{abe2018fashion,Kiapour2014,liu2016deepfashion,simo2015neuroaesthetics,takagi2017makes,yamaguchi2012parsing}. Table \ref{table:table1} presents the best-known public databases from fashion studies in computer science. 

Unfortunately, the fashion databases proposed in previous studies have several limitations in terms of the approach to everyday fashion trends. First, the level of detail of location annotation in existing databases is insufficient to analyze everyday fashion trends. People belong to a social community, and some social communities have their own distinct fashion styles and their territory. For instance, ‘Shibuya Fashion’ is a fashion style for young ladies that originated from a famous fashion mall in Shibuya~\cite{kawamura2006japanese}. Young ladies dressed in ‘Shibuya Fashion’ frequent Shibuya, one of the most famous fashion-conscious streets in Japan. Hence, we need fashion image data with location annotations at the street level to focus on what people wear in their daily lives.

Second, most databases consist of recent fashion images from the last decade because they were obtained from the internet. The periods covered by these databases might not be sufficiently long to determine fashion trends over longer periods. By examining how fashion changes over extended periods, sociologists and anthropologists have found that fashion has decadal-to-centennial trends and cyclic patterns~\cite{kroeber1919principle}. An example is the hemline index, which is a well-known hypothesis that describes the cyclic pattern in which skirt lengths decrease when economic conditions improve~\cite{Dhanorkar2015}. This pattern was determined based on observations of skirt lengths in the 1920s and 30s. 

Furthermore, we need data on how and what consumers wear to express their fashion styles. However, fashion photographs on the internet are one of two types: one displays clothes that consumers themselves choose to wear, and the other consists of photographs taken by professional photographers to promote a clothing line. As previous studies built databases by collecting fashion images from the internet, existing databases do not reflect only the fashion styles chosen by consumers. This is the third limitation of existing fashion data archives.

Taken together, no existing database has everyday fashion images that span over a long period with both timestamp and location information. In this study, we define everyday fashion trends as trends of fashion styles that consumers adopt in their daily lives. Everyday fashion exists on the streets~\cite{kawamura2006japanese,Rocamora2008}, changes over time, and trends and cyclic patterns can be observed over extended periods. Hence, we create a new fashion database, CAT STREET, is an attempt to solve these limitations and analyze everyday fashion trends.

\begin{table*}[tbp]
   \caption{Popular fashion databases used in computer science.}
   \begin{center}
      \begin{tabular}{ccccc}
         \hline
         Database name & Num. of images &\begin{tabular}{c} Geographical\\ information\end{tabular}&Time stamp                                          &\begin{tabular}{c}Fashion\\style tag\end{tabular}\\
         \hline \hline
         Fashionista~\cite{yamaguchi2012parsing}          &158,235           &                          &                                                     &\\
         Hipster Wars~\cite{Kiapour2014}           &1,893             &                          &                                                     &\checkmark\\
         DeepFashion~\cite{liu2016deepfashion}                &800,000           &	                       &                                                     &\checkmark\\
         Fashion 144k~\cite{simo2015neuroaesthetics}	       &144,169	        &City unit                 &		                                               &\checkmark\\
         FashionStyle14~\cite{takagi2017makes}          &13,126	           &		                    &                                                     &\checkmark\\
         When Was That Made?~\cite{vittayakorn2017made}&100,000           &		                    &\begin{tabular}{c}1900--2009\\Decade unit\end{tabular}&\\
         Fashion Culture Database~\cite{abe2018fashion}   &76,532,219        &City unit                 &\begin{tabular}{c}2000--2015\\Date unit\end{tabular}  &\\
         CAT STREET (Our Database)	             &14,688	           &Street unit               &\begin{tabular}{c}1970--2017\\Date unit\end{tabular}  &\\
         \hline
      \end{tabular}
   \end{center}
   \label{table:table1}
\end{table*}

\section{CAT STREET: Chronicle Archive of Tokyo Street-fashion}
We created CAT STREET via the following steps. We collected street-fashion photographs once or twice a month in fashion-conscious streets such as Harajuku and Shibuya in Tokyo from 1980 to date. In addition, we used fashion photographs from a third-party organization taken in the 1970s at monthly intervals in the same fashion-conscious streets. Next, by using images from the two data sources, we constructed a primary image database with timestamps from 1970 to 2017 for each image. The photographs from the third-party organization did not have location information; hence, we could only annotate the fashion images taken since 1980 with street tags.

Fashion styles conventionally differ between men and women. To focus on women’s fashion trends in CAT STREET, two researchers manually categorized the images by the subjects’ gender, reciprocally validating one another’s categorizations, and we selected only images of women from the primary image database. Some images from the 1970s were in monochrome; therefore, we gray-scaled all images to align the color tone across all images over the entire period.

\begin{figure*}[tbp]
   \begin{center}
     \includegraphics[width=0.93\linewidth]{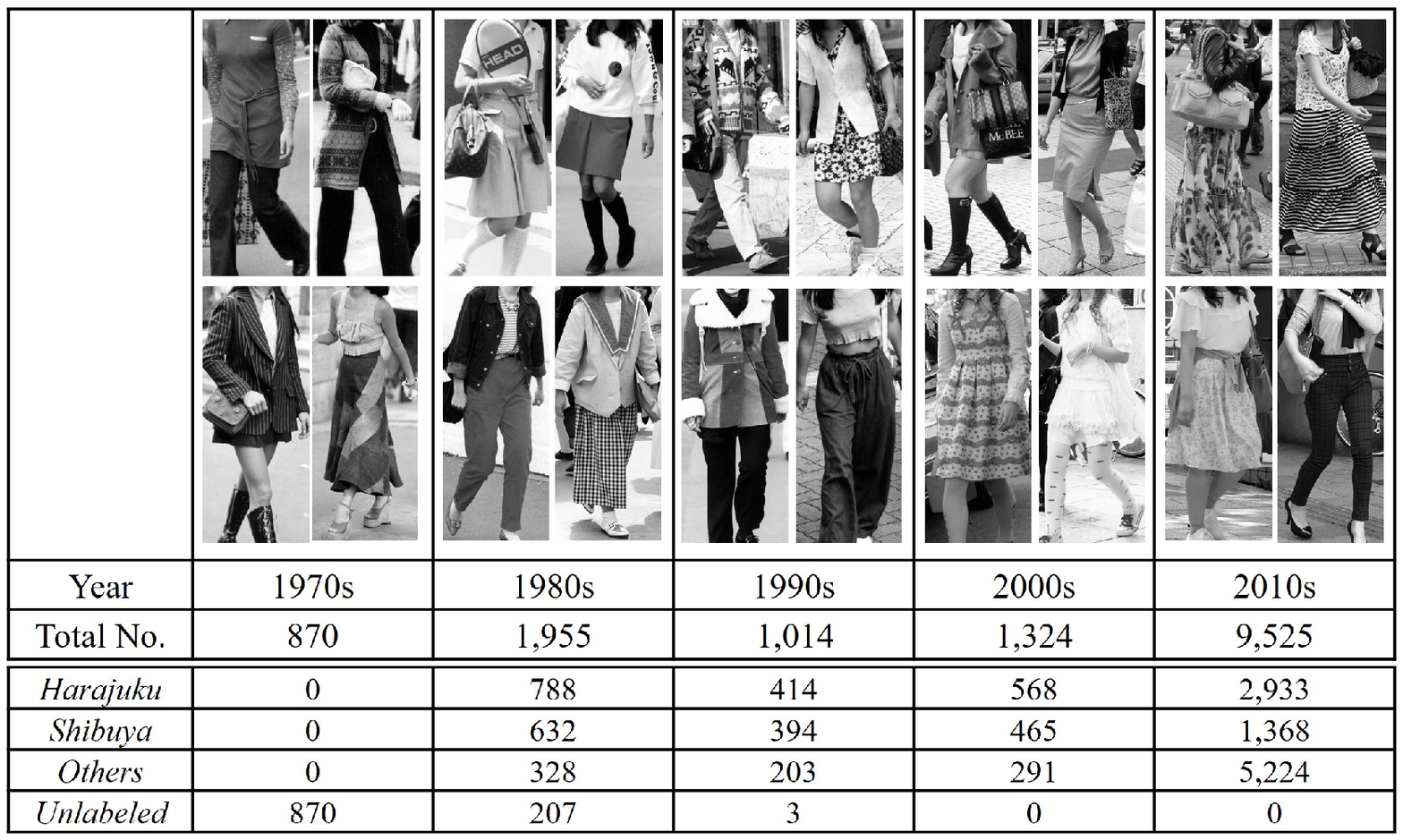}
   \end{center}
   \caption{Data overview of CAT STREET. Harajuku and Shibuya are the most famous streets in Japan and Others include images taken in other fashion-conscious streets such as  Ginza and Daikanyama.}
   \label{fig:fig6}
\end{figure*}

Street-fashion photographs generally contain a large amount of noise, which hinders the precise detection of combinations of clothes worn by the subjects of photographs. To remove the noise, we performed image pre-processing as the final step in the following manner. We identified human bodies in the photographs using OpenPose~\cite{Zhe2019openpose}, an machine learning algorithm for object recognition, and removed as much of the background image as possible. Subsequently, we trimmed the subject’s head to focus on clothing items based on the head position, which was detected using OpenPose. Fig. \ref{fig:fig6} shows an overview of the data contained in CAT STREET. The total number of images in the database was 14,688.

At the end of the database creation process, we checked whether CAT STREET met the requirements for a database to capture everyday fashion trends. First, CAT STREET comprises photographs captured on fashion-conscious streets in Tokyo. It reflects the fashion styles women wore in their real lives and does not include commercial fashion images produced for business purposes. Second, CAT STREET has necessary and sufficient annotations to track fashion trends: monthly timestamps from 1970 to 2017 and street-level location tags. Images in the 1970s do not have street tags; however, they were taken in the same streets as the photographs taken with street tags since 1980. Hence, we could use all the images in CAT STREET to analyze the overall trends of everyday fashion in fashion-conscious streets as a representative case in Japan.

\section{Retrieving Everyday Fashion Trends}
In this section, we estimated a fashion style clustering model to identify the fashion styles adopted in fashion images. Then, we applied the fashion clustering model to CAT STREET and retrieved everyday fashion trends indicating the extent to which people adopted the fashion styles in each year. Finally, we verified the robustness of clustering model's result.

\subsection{Fashion Style Clustering Model}

\begin{figure*}[tbp]
   \begin{center}
     \includegraphics[width=0.93\linewidth]{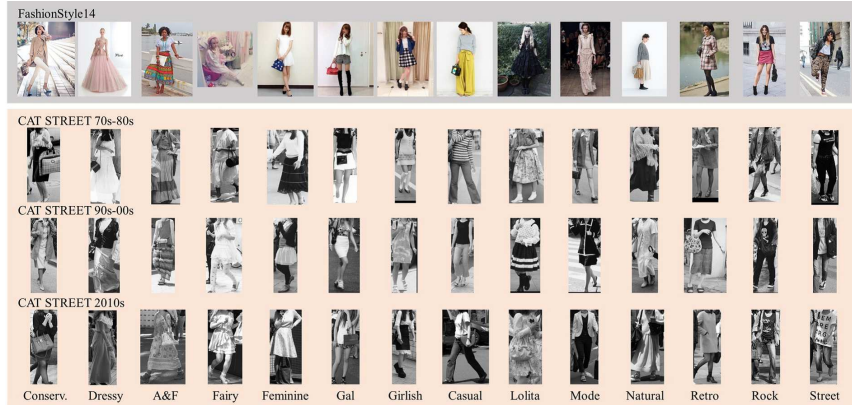}
   \end{center}
   \caption{Overview of FashionStyle14~\cite{takagi2017makes} and sample images in CAT STREET classified into different fashion styles using the fashion style clustering model. `Conserv.' is an abbreviation for Conservative. The A\&F is originally labeled as Ethnic~\cite{takagi2017makes}. We changed this original label to ‘A\&F’
because this category contains multiple fashion styles such as Asian, African, south-American, and Folklore styles.}
   \label{fig:fig7}
\end{figure*}

To build a fashion style clustering model, we selected FashionStyle14~\cite{takagi2017makes} as the training dataset. It consists of fourteen fashion style classes and their tags, as shown in the first row of Fig. \ref{fig:fig7}. Each class consists of approximately 1,000 images, and the database consists of a total of 13,126 images. The fashion styles of FashionStyle14 were selected by an expert as being representative of modern fashion trends in 2017. By applying the fashion clustering model to CAT STREET, we measured the share of each modern style in each year and regarded it as the
style’s trend.

We trained four deep learning network structures as options for our fashion clustering model: InceptionResNetV2~\cite{szegedy2017inception}, Xception~\cite{chollet2017xception}, ResNet50~\cite{he2016deep}, and VGG19~\cite{Simonyan2015}. We set weights trained on ImageNet~\cite{russakovsky2015imagenet} as the initial weights and fine-tuned them on FashionStyle14, which we gray-scaled to align the color tone of CAT STREET, using the stochastic gradient descent algorithm at a learning rate of of $10^{4}$. For fine-tuning, we applied k-fold cross-validation with k set as five.

The F1-scores are presented in Table \ref{table:table2}. InceptionResNetV2 yielded the highest F1-scores among the deep learning network structures. Its accuracy was 0.787, which is higher than the benchmark accuracy of 0.72 established by ResNet50 trained on FashionStyle14 in the study by Takagi et al.~\cite{takagi2017makes}. Therefore, we concluded that InceptionResNetV2 could classify gray-scaled color images to fashion styles and adopted the deep learning network structure InceptionResNetV2 as the fashion style clustering model in this study. We applied this fashion style clustering model to the images in CAT STREET, and Fig. \ref{fig:fig7} shows sample images classified into each fashion style.

\begin{table*}[bp]
   \caption{F1-scores of fine-tuned network architectures. The highest scores are underlined and in boldface.}
   \begin{center}
      \begin{tabular}{ccccc}
         \hline
                           &InceptionResNetV2   &Xception            &ResNet50&VGG19\\
         \hline
         Macro Avg.	      &\bf\underline{0.787}&0.782   	&0.747	&0.690\\
         Weighted Avg.	   &\bf\underline{0.786}&0.781   	&0.747	&0.689\\
         \hline
      \end{tabular}
   \end{center}
   \label{table:table2}
\end{table*}

\subsection{Verification}

The fashion style clustering model consisted of five models as we performed five-fold cross-validations when the deep learning network structure was trained, and each model estimated the style prevalence share for each image. To verify the clustering model’s robustness in terms of reproducing style shares, we evaluated the time-series correlations among the five models. Most fashion styles had high time-series correlation coefficients of over 0.8, and the unbiased standard errors were small. Some fashion styles, such as those designated Dressy and Feminine styles, exhibited low correlations because these styles originally had low style shares. These results indicate that our fashion style clustering model is a robust instrument for reproducing the time-series patterns of style shares. There were no images for 1997 and 2009; we replaced the corresponding zeros with the averages of the adjacent values for the analysis in the next section.

\section{Analyzing Fashion Trend Patterns}

There are two representative fashion-conscious streets in Tokyo: Harajuku and Shibuya. Geographically, Harajuku and Shibuya are very close and only a single transit station away from each other; however, they have different cultures. In particular, the everyday fashion trends in these streets are famously compared to each other. Some qualitative studies pointed out triggers that form the cultural and fashion modes in each street using an observational method~\cite{Hasegawa2015,kawamura2006japanese}. However, they simply reported the triggers with respect to the street and style and did not compare the functioning of these triggers on the mode formations in everyday fashion trends from a macro perspective. This section sheds light on this research gap and interprets how the fashion trends emerged with non-fashion data sources.

\begin{figure*}[bp]
   \begin{center}
     \includegraphics[width=0.93\linewidth]{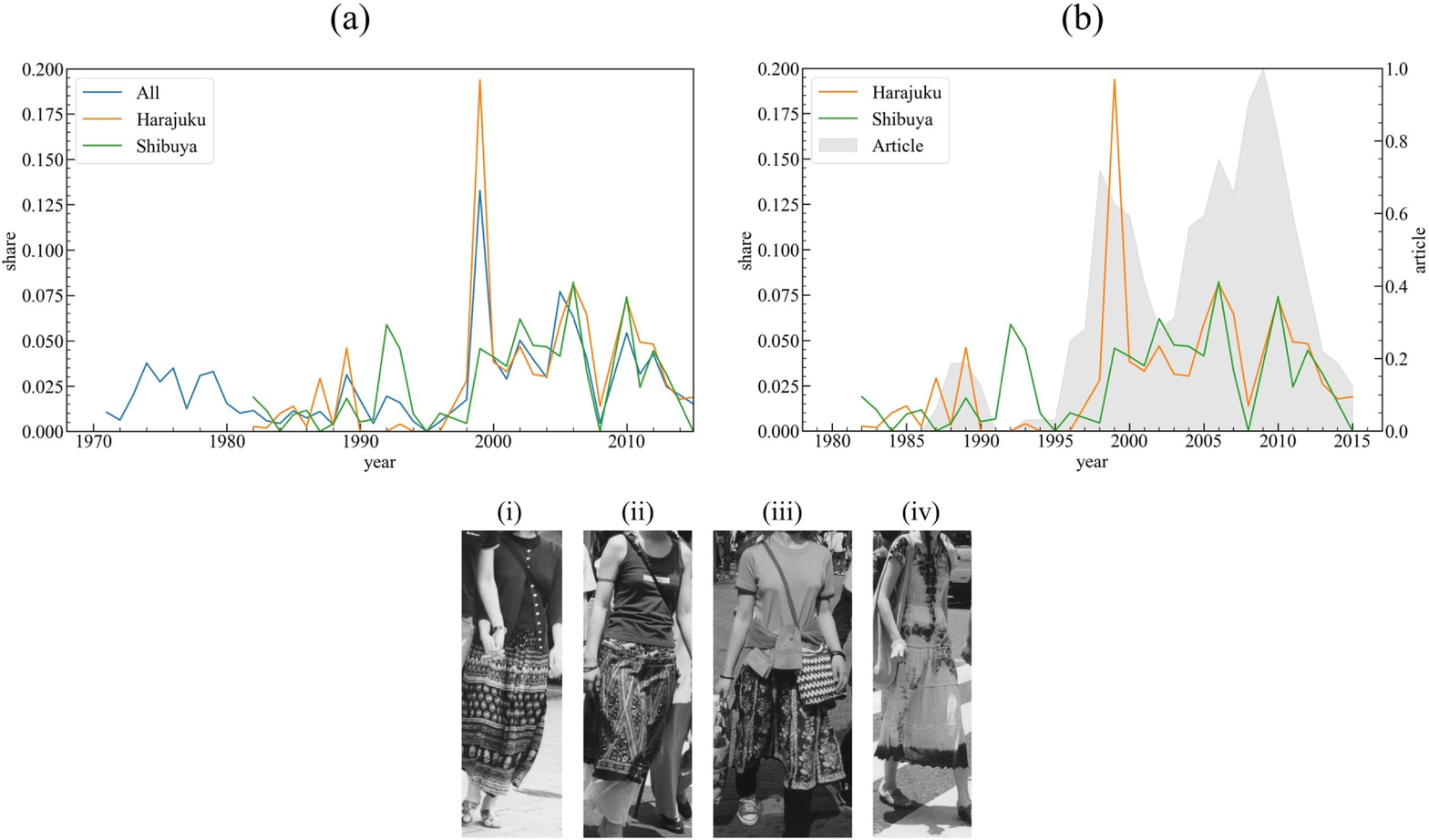}
   \end{center}
   \caption{A\&F style’s simultaneous emergence in both streets. (a) Average A\&F style share. (b) Number of articles about the A\&F style. Four images taken in Harajuku (i, ii) and Shibuya (iii, iv) in the late 1990s.}
   \label{fig:fig11}
\end{figure*}

\begin{figure*}[tbp]
   \begin{center}
     \includegraphics[width=0.93\linewidth]{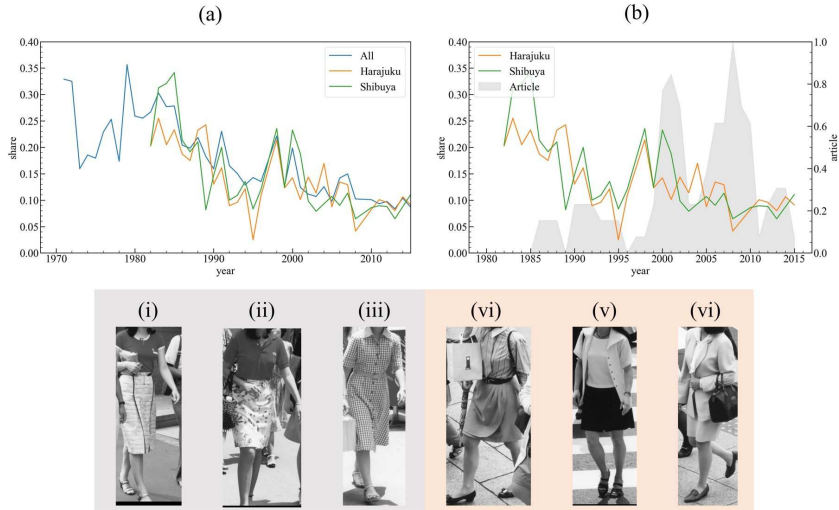}
   \end{center}
   \caption{Retro style’s simultaneous emergence in both streets. (a) Average Retro style share. (b) Number of articles about the 80s look. The three images on the left (i--iii) are examples of the Retro style taken in the early 1980s, and the three images on the right (iv--vi) are examples of the Retro style that came into fashion in the late 1990s.}
   \label{fig:fig12}
\end{figure*}

\begin{figure*}[tbp]
   \begin{center}
     \includegraphics[width=0.93\linewidth]{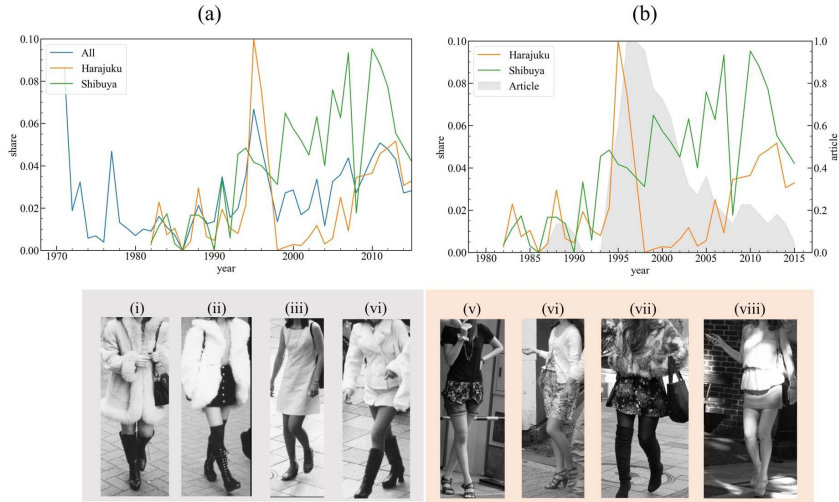}
   \end{center}
   \caption{Gal style’s trends in the streets showing different modes. (a) Average Gal style share. (b) Number of articles about the Gal style. The four images on the left were taken in Harajuku (i, ii) and Shibuya (iii, iv) in 1995--1996, and the four images on the right were taken in Harajuku (v, vi) and Shibuya (vii, viii) in the late 2000s.}
   \label{fig:fig13}
\end{figure*}

\begin{figure*}[tbp]
   \begin{center}
     \includegraphics[width=0.93\linewidth]{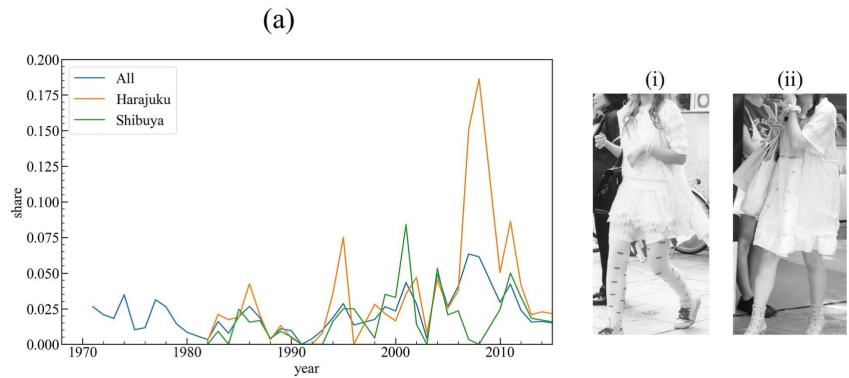}
   \end{center}
   \caption{Fairy style’s trends in the streets showing different modes. (a) Average Fairy style share. The two images (i and ii) were taken in Harajuku in the late 2000s.}
   \label{fig:fig14}
\end{figure*}

First, we compared two everyday fashion trends in fashion-conscious streets with CAT STREET to investigate how fashion styles ‘boom’ or suddenly increase in prevalence, and classified the resultant patterns. Fourteen styles were classified into two groups: a group comprising styles that emerged on both streets simultaneously and a group with differing timings for trends observed in the streets. We selected two fashion styles from each group as examples, including A\&F and Retro from the first group and Fairy and Gal from the second group; Figs. \ref{fig:fig11}(a), \ref{fig:fig12}(a), \ref{fig:fig13}(a), and \ref{fig:fig14}(a) show their average style shares, respectively.

\begin{figure*}[t]
   \begin{center}
     \includegraphics[width=0.93\linewidth]{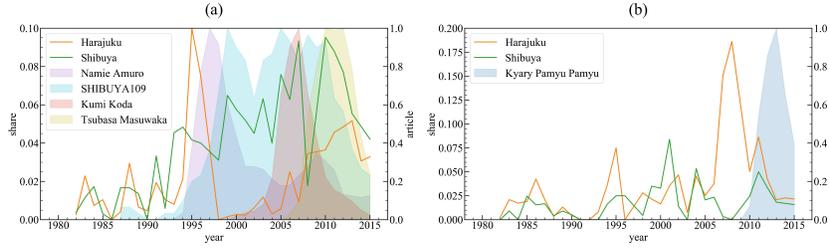}
   \end{center}
   \caption{(a) Gal style's share and 'magazine' trends for its icons. (b) Fairy style's share and 'magazine' trends for its icon.}
   \label{fig:fig15-6}
\end{figure*}

\begin{figure*}[t]
   \begin{center}
     \includegraphics[width=0.45\linewidth]{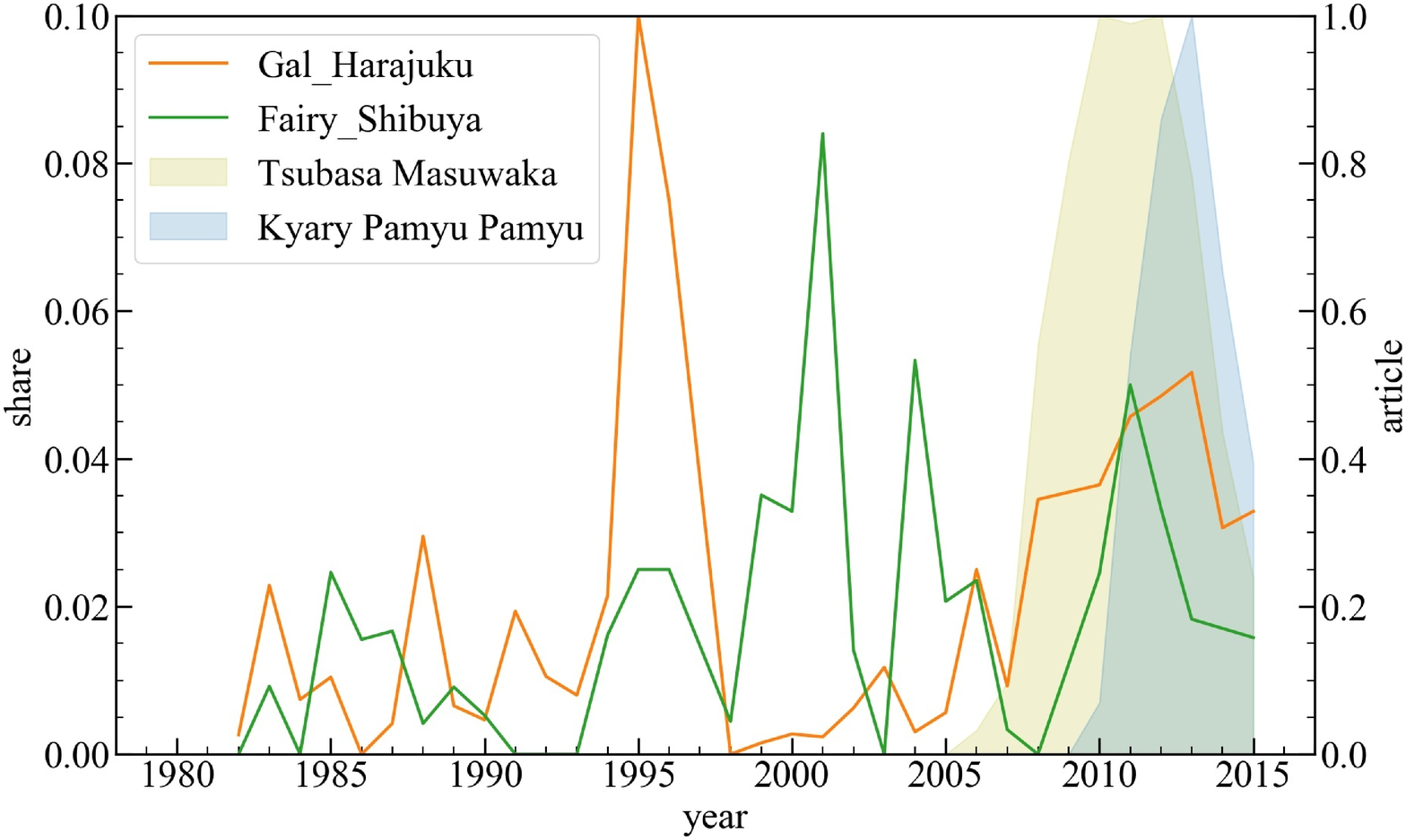}
   \end{center}
   \caption{Gal and Fairy style shares and 'magazine' trends for their icons.}
   \label{fig:fig17}
\end{figure*}
As a non-fashion data source, we focused our attention on magazines because fashion brands have built good partnerships with the magazine industry for a long time. Fashion brands regard magazines as essential media to build a bridge between themselves and consumers, and magazines play a role as a dispatcher of fashion in the market. We quantified the ‘magazine’ trends to capture how magazines or media sent out information to consumers. For this purpose, we used the digital magazine archive of Oya Soichi Library~\cite{OYA}. Oya Soichi Library has archived Japanese magazines since the late 1980s and built the digital archive. The digital archive houses about 1,500 magazines, and one can search for article headlines of about 4.5 million articles. 

The archive covers a range of age groups and mass-circulation magazines that people can easily acquire at small bookstores, convenience stores, and kiosks. By searching for headlines including fashion style words and specific topics from articles with tags related to fashion, we could quantify the ‘magazine’ trends that indicate how many articles dealt with the styles and specific topics to spread the information to a mass audience (Figs. \ref{fig:fig11}(b), \ref{fig:fig12}(b), \ref{fig:fig13}(b), \ref{fig:fig15-6}, and \ref{fig:fig17}). The ‘magazine’ trends labeled as Articles in the figures were normalized to the range of 0 to 1 for comparison.

Finally, to infer why and how the everyday fashion trends emerged, we compared the fashion style shares extracted from CAT STREET to the ‘magazine’ trends in the Oya Soichi Library database in chronological order.

\subsection{Simultaneous Emergence of Fashion Styles}
\label{Simultaneous Emergence of Fashion Styles}
We selected two styles, A\&F and Retro, as examples in the group of styles that simultaneously emerged on both streets. The A\&F style is inspired by native costumes~\cite{bunka1993}. Fig. \ref{fig:fig11}(a) shows the A\&F style’s upward trend in the late 1990s in Harajuku. Simultaneously, the A\&F style gradually became accepted in Shibuya in the late 1990s and reached the same share level in the mid-2000s. The Retro style, another example in the first group, is an abbreviation of retrospective style~\cite{bunka1993,Yoshimura2019}. According to this style’s definition, an overall downward trend is plausible in both streets, as shown in Fig. \ref{fig:fig12}(a). Fashion revival is one of the relevant fashion trend phenomena of the Retro style, and there are a wide variety of substyles representing fashion revivals under the Retro style, such as the 60s look, 70s look, and 80s look. In Fig. \ref{fig:fig12}(a), slight peaks can be observed in the early 1980s and late 1990s, suggesting that the 80s look, which was in vogue in the early 1980s, was revived in the late 1990s.

To quantify how magazines spread information about these styles, we plotted the ‘magazine’ trends by searching for articles with headlines that included terms related to the style names. Figs. \ref{fig:fig11}(b) and \ref{fig:fig12}(b) indicate two relationship patterns between style shares and ‘magazine’ trends; Fig. \ref{fig:fig11}(b) shows a synchronization phenomenon between style shares and magazine trends, while Fig. \ref{fig:fig12}(b) shows that magazine trends follow share trends.

We interpreted the first pattern as ‘mixed’; the shape of ‘magazine’ trends roughly matches that of style share trends, and this pattern is observed in the A\&F style. The ‘mixed’ patterns in the trends suggest that they include two types of phenomena between consumers and magazines. One is that the articles create new trends, spread them, and lead consumers to follow them. The other is that consumers create new style trends by themselves, and articles catch up on the new trends. These two types of phenomena could occur simultaneously or alternately. On the other hand, the second pattern is a ‘follow-up’; the articles dealt with the modes that were already in vogue on the streets and contributed to keeping their momentum for some time by spreading the information, and this pattern is observed in the Retro style during the 2000s. 

In this analysis, we found two relationship patterns between style shares and ‘magazine’ trends: ‘mixed’ and ‘follow-up’ patterns. We also inferred what types of interactions occur between consumers and magazines in each pattern. However, we could not perform an in-depth analysis on how to distinguish between the two types of consumer-magazine interactions in the case of the ‘mixed’ pattern because ‘magazine’ trends generated by searching for article headlines including terms related to the style names did not always reflect the contents of articles accurately. We also found it difficult to uncover why the article peak in the Retro style during the late 2000s was seemingly irrelevant to the Retro style share. To approach these unsolved research questions as future work, we must perform text mining on article contents and headlines, and categorize the articles into the creative type, which creates and spreads new trends, or the reporting type, which refers to presently existing new consumer trends after they develop; e.g. ‘ten trends coming in autumn, checker-board pattern, check pattern, military, big tops, shoes \& boots, foot coordination, fur, A\&F vs. Nordic, beautiful romper, best of the season’ (translated from Japanese by the authors) is categorized as a creative-type article title, whereas ‘A\&F style is a hot topic this spring. We will introduce you to some of the Japanese items that are getting a lot of attention. Crepe, goldfish prints, and more...’ (translated from Japanese by the authors) is categorized as a reporting-type article title. By measuring the number of articles about the fashion styles and identifying the contents of articles to reflect the types of consumer-magazine relationships, we expect to decompose relationship types in the ‘mixed’ pattern, interpret seemingly irrelevant relationships, and clarify the role of each relationship type in the trend formation process.

\subsection{Emergence of Fashion Styles at Different Types in the Streets}
\label{Emergence of Fashion Styles at Different Types in the Streets}
For the case where fashion trends are observed at different times in the two streets, we focused on two styles: the Gal style, which can be characterized as an exaggeration of the American teenage party style, and the Fairy style, which involves the fashion coordination of frilly dresses reminiscent of fairies~\cite{Yoshimura2019}.

Figs. \ref{fig:fig13} and \ref{fig:fig14} show how the acceptances of these two styles changed in each street. The Gal style came into fashion around 1995 simultaneously in Harajuku and Shibuya. The style remained in Shibuya, whereas it lost its popularity quickly in Harajuku but re-emerged in the late 2000s. The Fairy style emerged in the late 2000s in Harajuku only and lost its momentum in the early 2010s. As with the first case, we searched ‘magazine’ trends for these style names and compared them to the style shares. For the Gal style, we found that the style and ‘magazine’ trends correlated with each other in the mid-80s and the mid-90s. However, magazines gradually lost interest in the style after the mid-00s, and the relationship between the style and ‘magazine’ trend disappeared accordingly. For the Fairy style, on the other hand, no article in the digital magazine archive included the style name in the headlines. This is because ‘Fairy’ is a kind of jargon among people interested in the style and hence not used in magazines.

What accounts for this difference between the first and the second case? We assumed that the critical factor determining people’s choice in fashion, i.e. ‘why people adopt a given style,’ is for some purpose; some styles have specific features, such as colors, silhouettes, and patterns, that people consume as a fashion or which represent a certain social identity and have some characteristic features. The Gal and Fairy styles are in the latter group and represent a ‘way of life’ for some people~\cite{Hasegawa2015,Kyary2011}.

Icons representing ‘way of life,’ such as celebrities, have substantial power to influence people to behave in a certain manner; for instance, celebrities can influence people to purchase products that they use or promote. Choosing or adopting fashion styles is no exception, and some articles identified fashion icons for the Gal and the Fairy styles~\cite{Hasegawa2015,Shinmura2012,Yoshimura2019}. We attempted to explain why the style trends in the streets showed different modes from the perspective of fashion icons and the media type that the icons utilized.

Figs. \ref{fig:fig15-6} and \ref{fig:fig17} show the relationship between the style shares and ‘magazine’ trends for style icons. Previous works introduced the following Gal style icons: Namie Amuro for the mid-90s, SHIBUYA 109 for the late-00s, Ayumi Hamasaki for the early-00s, and Kumi Koda and Tsubasa Masuwaka for the late-00s~\cite{Hasegawa2015,Shinmura2012}.

The first icon who boosted the Gal style was Namie Amuro, who debuted nationwide as a pop singer in 1992. Girls yearned to imitate her fashion style, and magazines focused heavily on her as a style icon in the mid-90s. Simultaneously, as the ‘magazine’ trend for the style shows in Fig. \ref{fig:fig15-6} (a), many tabloid magazines were interested in this phenomenon and spread it as a new ‘way of life’ among youth consumers. The first icon boosted the style in both streets because she appeared nation-wide. However, the second icon that sustained the Gal style’s upward trend is regionally specific: an iconic fashion mall named SHIBUYA 109 (pronounced Ichi-maru-kyū) in Shibuya. Many tenants in the mall sold Gal style products, which is why the Gal style is also known as ‘maru-kyū fashion’~\cite{Yoshimura2019}. This street-specific image created by the second icon might have influenced the third and fourth icons, Ayumi Hamasaki and Kumi Koda, who were nationwide pop singers in the early-00s and the mid-00s, because the style share increased during that time only in Shibuya.

The icons prompted the last increase in the prevalence share of the Gal style in Shibuya, and the spike in the Fairy style’s share in Harajuku exhibited the same characteristics. Both Tsubasa Masuwaka for the Gal style in Shibuya and ‘Kyary Pamyu Pamyu’ for the Fairy style in Harajuku were active as exclusive reader models in street-based magazines at the beginning, and they created the booms in each street~\cite{Shinmura2012}. However, the street-based magazines they belonged to were not included in the digital magazine archive in Oya Soichi Library; hence, the ‘magazine’ trends for their names missed their activities as exclusive reader models when the style shares showed an upward trend (Figs. \ref{fig:fig15-6} (b) and \ref{fig:fig17}). Around 2010, they debuted as nationwide pop stars and frequently appeared in mass-circulation magazines and on television. Additionally, Tsubasa Masuwaka proposed a new style named ‘Shibu-Hara’ fashion, a combination of the styles in both Shibuya and Harajuku, and also referred to interactions with ‘Kyary Pamyu Pamyu’ on her social networking account. The Gal style’s second peak in Harajuku and the Fairy style’s spike in Shibuya in the early 2010s aligned with these icons’ nationwide activities; this indicates the styles’ cross-interactions driven by the icons (Figs. \ref{fig:fig15-6} (b) and \ref{fig:fig17}).

\subsection{Discussion}
The findings in Section \ref{Simultaneous Emergence of Fashion Styles} prompted a new research question of determining the importance of the different roles that media play in fashion trends: the role of the ‘mixed’ pattern for creating and reporting new trends and the ‘follow-up’ effect to keep their momentum. To theorize how fashion trends are generated in more detail, our findings indicate that quantifying the trends from multiple digital archives is essential to test the research questions about the effect of the media’s role, which has not been discussed thoroughly.

In previous studies, researchers analyzed fashion style trends at a street level independently. However, to analyze the recent, more complicated trend patterns, the findings in Section \ref{Emergence of Fashion Styles at Different Types in the Streets} suggests that it would be thought-provoking to focus on what the icons represent for people’s social identities, rather than the style itself, and how they lead the trends. An example is the analysis of the types of media that icons use and the reach of that media. The viewpoints gained from our findings can be useful in tackling unsolved research questions such as how styles interact and how new styles are generated from this process. If we can determine people’s social identities from the aims of their adopted fashions using the digital archives, researchers can retrieve beneficial information to delve into the research questions and expand fashion trend theories.

\section{Conclusions}
In this study, we reviewed the issues in the existing digital fashion archives for the analysis of everyday fashion trends. As one of the solutions to these issues, we built CAT STREET, which comprises fashion images illustrating what people wore in their daily lives in the period 1970--2017, along with street-level geographical information. We demonstrated that machine learning retrieved how fashion styles emerged on the street using CAT STREET and the magazine archives helped us interpret these phenomena. These empirical analyses showed our approach’s potential to find new perspectives to promote the understanding of our societies and cultures through fashion embedded in the daily lives of consumers.

Our work is not without limitations. We used the fashion style categories of FashionStyle14~\cite{takagi2017makes}, which was defined by fashion experts and is considered to represent modern fashion. However, the definition does not cover all contemporary fashion styles and their substyles in a mutually exclusive and collectively exhaustive manner. Defining fashion styles is a complicated task because some fashion styles emerge from consumers, and others are defined by suppliers. Hence, we must further refine the definition of fashion styles to capture everyday fashion trends more accurately. Furthermore, prior to building CAT STREET, only printed photos were available for the period 1970--2009. Consequently, the numbers of images for these decades are not equally distributed because only those images from printed photos that have already undergone digitization are currently present in the database. The remainder of the printed photos will be digitized and their corresponding images added to the database in future work.

%
%
%
\bibliographystyle{splncs04}
\bibliography{references}

\begin{thebibliography}{10}
\providecommand{\url}[1]{\texttt{#1}}
\providecommand{\urlprefix}{URL }
\providecommand{\doi}[1]{https://doi.org/#1}

\bibitem{abe2018fashion}
Abe, K., Minoguchi, M., Suzuki, T., Suzuki, T., Akimoto, N., Qiu, Y., Suzuki,
  R., Iwata, K., Satoh, Y., Kataoka, H.: {F}ashion {C}ulture {D}atabase:
  {C}onstruction of {D}atabase for {W}orld-wide {F}ashion {A}nalysis. In:
  Proceedings of the 2018 15th {I}nternational {C}onference on {C}ontrol,
  {A}utomation, {R}obotics and {V}ision {(ICARCV)}. pp. 1721--1726. IEEE (Nov
  2018). \doi{10.1109/ICARCV.2018.8581148}

\bibitem{belleau1987cyclical}
Belleau, B.D.: {Cyclical Fashion Movement: Women's Day Dresses: 1860-1980}.
  {Clothing and Textiles Research Journal}  \textbf{5}(2),  15--20 (1987).
  \doi{10.1177/0887302X8700500203}

\bibitem{bunka1993}
{Bunka Gakuen Bunka Publishing Bureau}, {Bunka Gakuen University Textbook
  Department} (eds.): Fashion Dictionary. {Bunka Gakuen Bunka Publishing
  Bureau}, Tokyo, Japan (1993)

\bibitem{Zhe2019openpose}
Cao, Z., Hidalgo, G., Simon, T., Wei, S.E., Sheikh, Y.: {OpenPose: Realtime
  Multi-Person 2D Pose Estimation Using Part Affinity Fields}. {IEEE
  Transactions on Pattern Analysis and Machine Intelligence}  \textbf{43}(1),
  172--186 (2019)

\bibitem{chollet2017xception}
Chollet, F.: {Xception: Deep Learning With Depthwise Separable Convolutions}.
  In: {Proceedings of the 2017 IEEE Conference on Computer Vision and Pattern
  Recognition (CVPR)}. pp. 1251--1258. IEEE (July 2017).
  \doi{10.1109/CVPR.2017.195}

\bibitem{Vogue}
{Condé Nast}: {Vogue Archive}. \url{https://archive.vogue.com/}, accessed:
  2021-07-11

\bibitem{Dhanorkar2015}
Dhanorkar, S.: 8 unusual indicators to gauge economic health around the world.
  The Economic Times  (2015)

\bibitem{EuropeanaFashion}
{Europeana}: {Europeana, Fashion}.
  \url{https://www.europeana.eu/en/collections/topic/ 55-fashion}, accessed:
  2021-07-11

\bibitem{Hasegawa2015}
Hasegawa, S.: {The History of Two Decades for Gals and ``Guys``
  1990's--2010's}. Akishobo Inc., Tokyo, Japan (2015)

\bibitem{he2016deep}
He, K., Zhang, X., Ren, S., Sun, J.: {Deep Residual Learning for Image
  Recognition}. In: {Proceedings of the 2016 IEEE Conference on Computer Vision
  and Pattern Recognition (CVPR)}. pp. 770--778. IEEE (Jun 2016).
  \doi{10.1109/CVPR.2016.90}

\bibitem{JapaneseFashionArchive}
japanese.fashion.archive: {Japanese Fashion Archive}.
  \url{http://japanesefashionarchive.com/about/}, accessed: 2021-07-11

\bibitem{kawamura2006japanese}
Kawamura, Y.: {Japanese Teens as Producers of Street Fashion}. {Current
  Sociology}  \textbf{54}(5),  784--801 (2006). \doi{10.1177/0011392106066816}

\bibitem{kestemont2017lemmatization}
Kestemont, M., De~Pauw, G., van Nie, R., Daelemans, W.: Lemmatization for
  variation-rich languages using deep learning. Digital Scholarship in the
  Humanities  \textbf{32}(4),  797--815 (2017). \doi{10.1093/llc/fqw034}

\bibitem{Kiapour2014}
Kiapour, M.H., Yamaguchi, K., Berg, A.C., Berg, T.L.: {Hipster Wars:
  Discovering Elements of Fashion Styles}. In: Fleet, D., Pajdla, T., Schiele,
  B., Tuytelaars, T. (eds.) Computer Vision -- ECCV 2014. pp. 472--488.
  Springer International Publishing, Cham (2014)

\bibitem{kroeber1919principle}
Kroeber, A.L.: {On the Principle of Order in Civilization as Exemplified by
  Changes of Fashion}. American Anthropologist  \textbf{21}(3),  235--263
  (1919)

\bibitem{Kyary2011}
{Kyary Pamyu Pamyu}: Oh! My God!! Harajuku Girl. POPLAR Publishing Co., Ltd.,,
  Tokyo, Japan (2011)

\bibitem{lancioni1973brief}
Lancioni, R.A.: A brief note on the history of fashion. Journal of the Academy
  of Marketing Science  \textbf{1},  128--131 (1973). \doi{10.1007/BF02722015}

\bibitem{lang2018attesting}
Lang, S., Ommer, B.: {Attesting similarity: Supporting the organization and
  study of art image collections with computer vision}. Digital Scholarship in
  the Humanities  \textbf{33}(4),  845--856 (2018). \doi{10.1093/llc/fqy006}

\bibitem{liu2016deepfashion}
Liu, Z., Luo, P., Qiu, S., Wang, X., Tang, X.: {DeepFashion: Powering Robust
  Clothes Recognition and Retrieval with Rich Annotations}. In: {Proceedings of
  the 2016 IEEE Conference on Computer Vision and Pattern Recognition (CVPR)}.
  pp. 1096--1104. IEEE (Jun 2016)

\bibitem{LACMAFashion}
{Los Angeles County Museum of Art}: {LACMA, Fashion, 1900-2000}.
  \url{https://collections.lacma.org/node/589000}, accessed: 2021-07-11

\bibitem{lowe1993quantitative}
Lowe, E.D.: {Quantitative Analysis of Fashion Change: A Critical Review}. Home
  Economics Research Journal  \textbf{21}(3),  280--306 (1993).
  \doi{10.1177/0046777493213004}

\bibitem{lowe1990velocity}
Lowe, E.D., Lowe, J.W.: {Velocity of the Fashion Process in Women's Formal
  Evening Dress, 1789-1980}. Clothing and Textiles Research Journal
  \textbf{9}(1),  50--58 (1990). \doi{10.1177/0887302X9000900107}

\bibitem{BostonMuseum}
{Museum of Fine Arts, Boston}: {Boston Museum of Fine Arts}.
  \url{https://collections.mfa.org/collections/419373/tfafashionable-dress;jsessionid=05B040E674A761851F36C9A1C70DBB7F/objects},
  accessed: 2021-07-11

\bibitem{OYA}
{OYA-bunko}: {Web OYA-bunko}. \url{https://www.oya-bunko.com/}, accessed:
  2021-07-11

\bibitem{robenstine1981relating}
Robenstine, C., Kelley, E.: {Relating Fashion Change to Social Change: A
  Methodological Approach}. Home Economics Research Journal  \textbf{10}(1),
  78--87 (1981). \doi{10.1177/1077727X8101000110}

\bibitem{Rocamora2008}
Rocamora, A., O'Neill, A.: {Fashioning the Street: Images of the Street in the
  Fashion Media}. In: Shinkle, E. (ed.) {Fashion as Photograph Viewing and
  Reviewing Images of Fashion}, chap.~13, pp. 185--199. I.B.Tauris, London
  (2008). \doi{10.5040/9780755696420.ch-013}

\bibitem{russakovsky2015imagenet}
Russakovsky, O., Deng, J., Su, H., Krause, J., Satheesh, S., Ma, S., Huang, Z.,
  Karpathy, A., Khosla, A., Bernstein, M., et~al.: {ImageNet Large Scale Visual
  Recognition Challenge}. International Journal of Computer Vision
  \textbf{115}(3),  211--252 (2015). \doi{doi.org/10.1007/s11263-015-0816-y}

\bibitem{Shinmura2012}
Shinmura, K., Okamoto, A.: {2012 The Frontline of the Never-Ending "Reader
  models" Boom}. Weekly Playboy (23),  89--100 (2012)

\bibitem{simo2015neuroaesthetics}
Simo-Serra, E., Fidler, S., Moreno-Noguer, F., Urtasun, R.: {Neuroaesthetics in
  fashion: Modeling the perception of fashionability}. In: Proceedings of the
  2015 IEEE Conference on Computer Vision and Pattern Recognition (CVPR). pp.
  869--877. IEEE (Jun 2015)

\bibitem{Simonyan2015}
Simonyan, K., Zisserman, A.: {Very Deep Convolutional Networks for Large-Scale
  Image Recognition}. In: Bengio, Y., LeCun, Y. (eds.) Proceedings of the 3rd
  International Conference on Learning Representations (ICLR) (May 2015)

\bibitem{szegedy2017inception}
Szegedy, C., Ioffe, S., Vanhoucke, V., Alemi, A.: {Inception-v4,
  inception-ResNet and the impact of residual connections on learning}. In:
  Proceedings of the Thirty-First AAAI Conference on Artificial Intelligence.
  pp. 4278--4284. AAAI Press (Feb 2017)

\bibitem{takagi2017makes}
Takagi, M., Simo-Serra, E., Iizuka, S., Ishikawa, H.: {What Makes a Style:
  Experimental Analysis of Fashion Prediction}. In: Proceedings of the 2017
  IEEE International Conference on Computer Vision Workshops (ICCVW). pp.
  2247--2253. IEEE (Oct 2017). \doi{10.1109/ICCVW.2017.263}

\bibitem{MET}
{The Metropolitan Museum of Art}: {THE MET 150, Libraries and Research Centers,
  ART, Thomas J. Watson Library Digital Collections, Costume Institute
  Collections}.
  \url{https://www.metmuseum.org/art/libraries-and-research-centers/watson-digital-collections/costume-institute-collections},
  accessed: 2021-07-11

\bibitem{TheMuseumatFIT}
{The Museum at FIT}: {The Museum at FIT, Collections}.
  \url{https://www.fitnyc.edu//museum/collections/}, accessed: 2021-07-11

\bibitem{UNT}
{University of North Texas}: {UNT Digital Library, Texas Fashion Collection}.
  \url{https://digital.library.unt.edu/explore/collections/TXFC/}, accessed:
  2021-07-11

\bibitem{vittayakorn2017made}
Vittayakorn, S., Berg, A.C., Berg, T.L.: When was that made? In: Proceedings of
  the 2017 IEEE Winter Conference on Applications of Computer Vision (WACV).
  pp. 715--724. IEEE (Mar 2017). \doi{10.1109/WACV.2017.85}

\bibitem{wevers2020visual}
Wevers, M., Smits, T.: The visual digital turn: Using neural networks to study
  historical images. Digital Scholarship in the Humanities  \textbf{35}(1),
  194--207 (2020). \doi{10.1093/llc/fqy085}

\bibitem{yamaguchi2012parsing}
Yamaguchi, K., Kiapour, M.H., Ortiz, L.E., Berg, T.L.: Parsing clothing in
  fashion photographs. In: Proceedings of the 2012 IEEE Conference on Computer
  Vision and Pattern Recognition (CVPR). pp. 3570--3577. IEEE (Jun 2012)

\bibitem{Yoshimura2019}
Yoshimura, S.: {The Fashion Logos}. Senken Shinbun Co., Ltd., Tokyo, Japan
  (2019)

\end{thebibliography}
\end{document}